\documentclass[onecolumn,floatfix]{article}

\usepackage{graphicx,color} 
\usepackage{array,hhline,dcolumn} 
\usepackage{rotating} 
\usepackage{verbatim}

\usepackage[english]{babel}
\usepackage{subfigure}
\usepackage{epsfig}
\usepackage{amsmath}
\usepackage{amssymb}
\usepackage{amsfonts}
\usepackage{slashed}
\usepackage{bbm}
\usepackage{cancel}
\usepackage{wrapfig}
\usepackage[labelfont=bf]{caption}
\usepackage[section]{placeins}
\usepackage{fullpage}

 \DeclareGraphicsExtensions{.eps,.ps,.eps.gz,.ps.gz,.pdf}    

\newcommand{\beq}{\begin{equation}}
\newcommand{\eeq}{\end{equation}}
\newcommand{\beqa}{\begin{eqnarray}}
\newcommand{\eeqa}{\end{eqnarray}}
\newcommand{\bitm}{\begin{itemize}}
\newcommand{\eitm}{\end{itemize}}
\newcommand{\btab}{\begin{table}}
\newcommand{\etab}{\end{table}}

\newcommand{\nue}{\nu_e}

\renewcommand{\nue}{\nu_e}

\newcommand{\gtwid}{\mathrel{\raise.3ex\hbox{$>$\kern-.75em\lower1ex\hbox{$\sim$}}}}
\newcommand{\ltwid}{\mathrel{\raise.3ex\hbox{$<$\kern-.75em\lower1ex\hbox{$\sim$}}}}

\begin{document}

\title{OscSNS: A Precision Neutrino Oscillation Experiment at the SNS}

\author{The OscSNS Collaboration}
\maketitle

\begin{abstract}
The growing evidence for short-baseline neutrino oscillations and the possible
existence of sterile neutrinos necessitates the development of a cost-effective experiment that 
can resolve these mysteries.  The OscSNS~\cite{1} experiment, 
located at the Spallation Neutron Source (SNS), Oak Ridge Laboratory, is ideal for this purpose. 
\end{abstract}

\vspace{1cm}

\begin{figure}
\centering
\includegraphics[scale=0.75,angle=0]{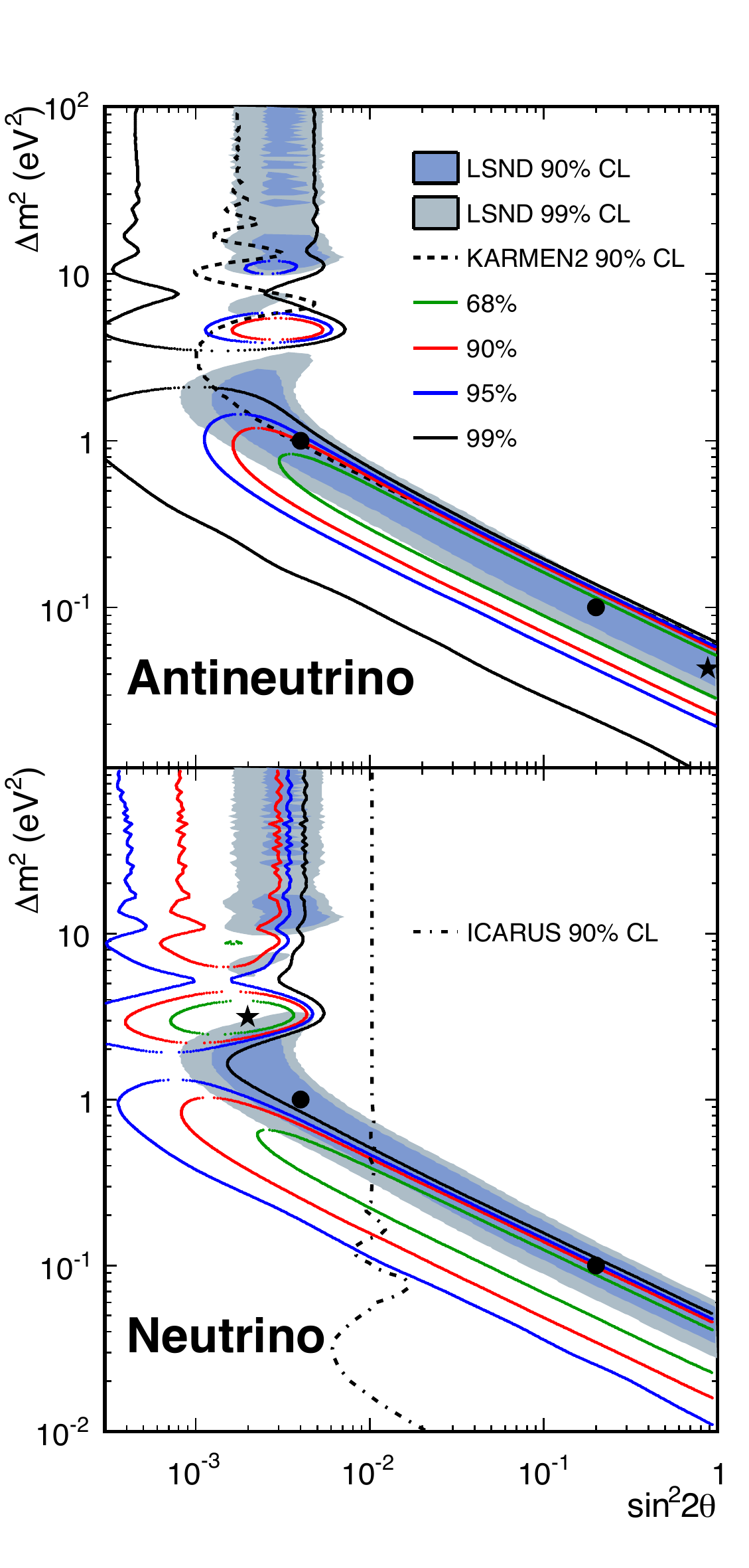}
\caption{
MiniBooNE allowed regions in antineutrino mode (top)
and neutrino mode (bottom) for events with
$E^{QE}_{\nu} > 200$ MeV within a two-neutrino oscillation model.
Also shown are the ICARUS~\cite{7} and KARMEN~\cite{8}
appearance limits for neutrinos and antineutrinos, respectively.
The shaded areas show the 90\% and 99\% C.L. LSND
$\bar{\nu}_{\mu}\rightarrow\bar{\nu}_e$ allowed
regions. The black stars show the MiniBooNE best fit points,
while the circles
show central values from LSND.
}
\label{mb_osc}
\end{figure}

There exists a need to address and resolve the growing evidence 
for short-baseline neutrino oscillations and the possible existence of sterile neutrinos.
Such non-standard particles, first invoked to explain the LSND $\bar \nu_\mu \rightarrow \bar \nu_e$ 
appearance signal~\cite{2}, would require a mass of $\sim 1$ eV/c$^2$,
far above the mass scale associated with active neutrinos. More recently, as shown
in the top plot of Figure~\ref{mb_osc}, the
MiniBooNE experiment has reported a $2.8 \sigma$ excess of events
in antineutrino mode that is consistent with neutrino oscillations and
with the LSND antineutrino appearance signal~\cite{3}. MiniBooNE has also observed a 
$3.4 \sigma$ excess of events in their neutrino mode data.  In addition, lower than
expected neutrino-induced event rates from calibrated radioactive sources~\cite{4} and
nuclear reactors~\cite{5} can also be explained by the existence of sterile neutrinos.
Fits to the world's neutrino and antineutrino data are consistent with
sterile neutrinos at this $\sim 1$ eV/c$^2$ mass scale, although there is some
tension between measurements from disappearance and appearance experiments~\cite{6,6a}.  The existence of 
these sterile neutrinos will impact design and planning 
for next generation neutrino experiments.  It should be conclusively established whether such totally unexpected light sterile neutrinos exist.
The Spallation Neutron Source (SNS) at Oak Ridge National Laboratory, 
built to usher in a new era in neutron research, provides a unique opportunity
for US science to perform a definitive search for sterile neutrinos.

The 1.4 MW beam power of the SNS is a prodigious source of neutrinos from the decay of $\pi^+$ 
and $\mu^+$ at rest. These decays produce a well specified flux of neutrinos via
$\pi^+ \rightarrow \mu^+$ $\nu_\mu$, $\tau_\pi = 2.7 \times 10^{-8}$ s, and $\mu^+ \rightarrow e^+$ 
 $\nu_e$ $\bar \nu_\mu$, $\tau_\mu = 2.2 \times 10^{-6}$ s.
The low duty factor of the SNS ($\sim 695$ ns beam pulses at 60 Hz, $DF=4.2 \times 10^{-5}$) is more than 
1000 times less than that found at LAMPF.  This smaller duty factor provides a reduction in backgrounds 
due to cosmic rays, and allows the $\nu_\mu$ induced events from $\pi^+$ 
decay to be separated from the $\nu_e$ and $\bar \nu_\mu$ induced events from $\mu^+$ decay.

\begin{figure}
\centering
\includegraphics[scale=0.50,angle=90]{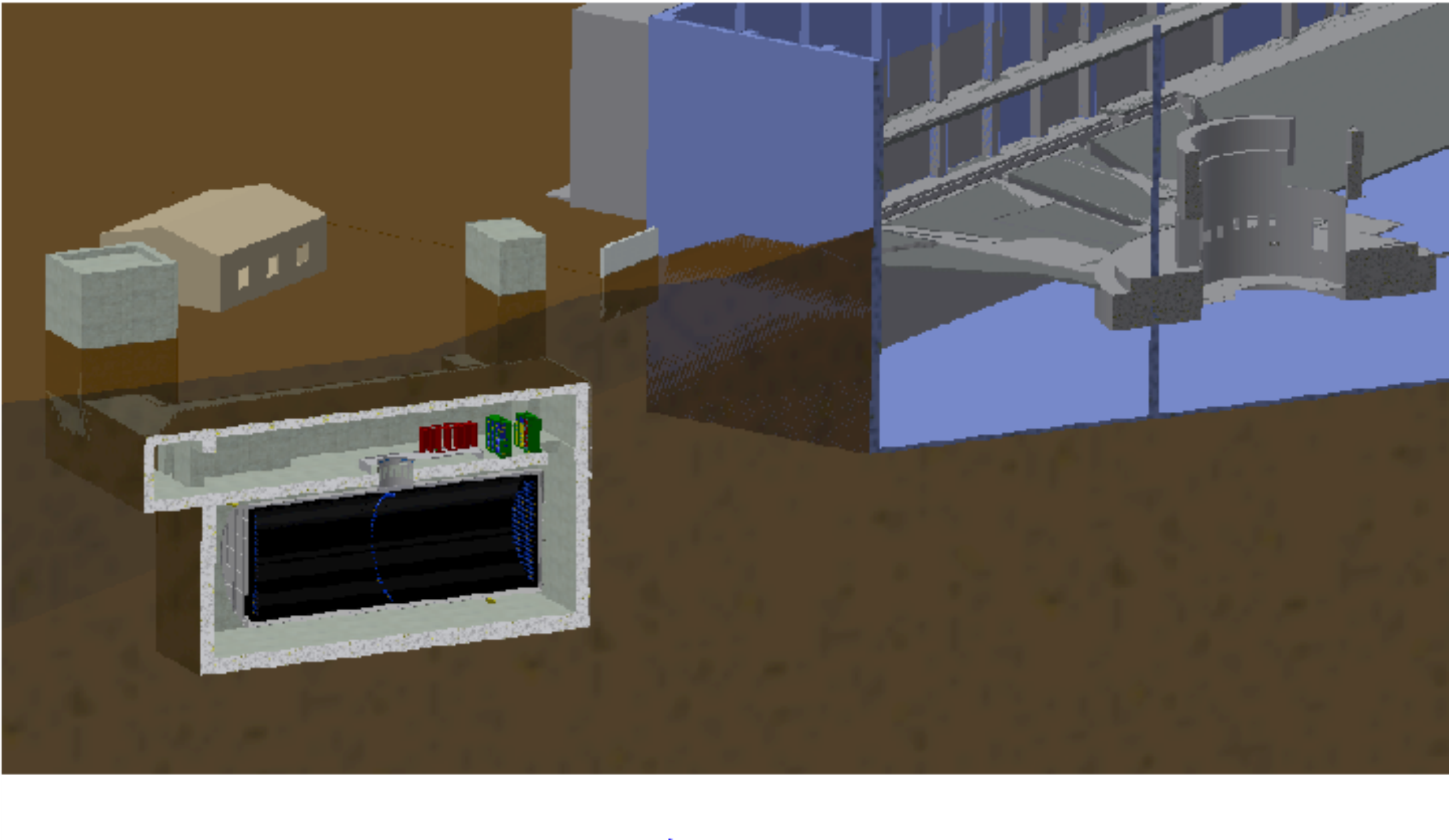}
   \caption{A schematic drawing of the suggested detector location in relation
to the SNS target hall.}
\label{detector_target_schematic1}
\end{figure}

\begin{figure}
\centering
\includegraphics[scale=0.4,angle=90]{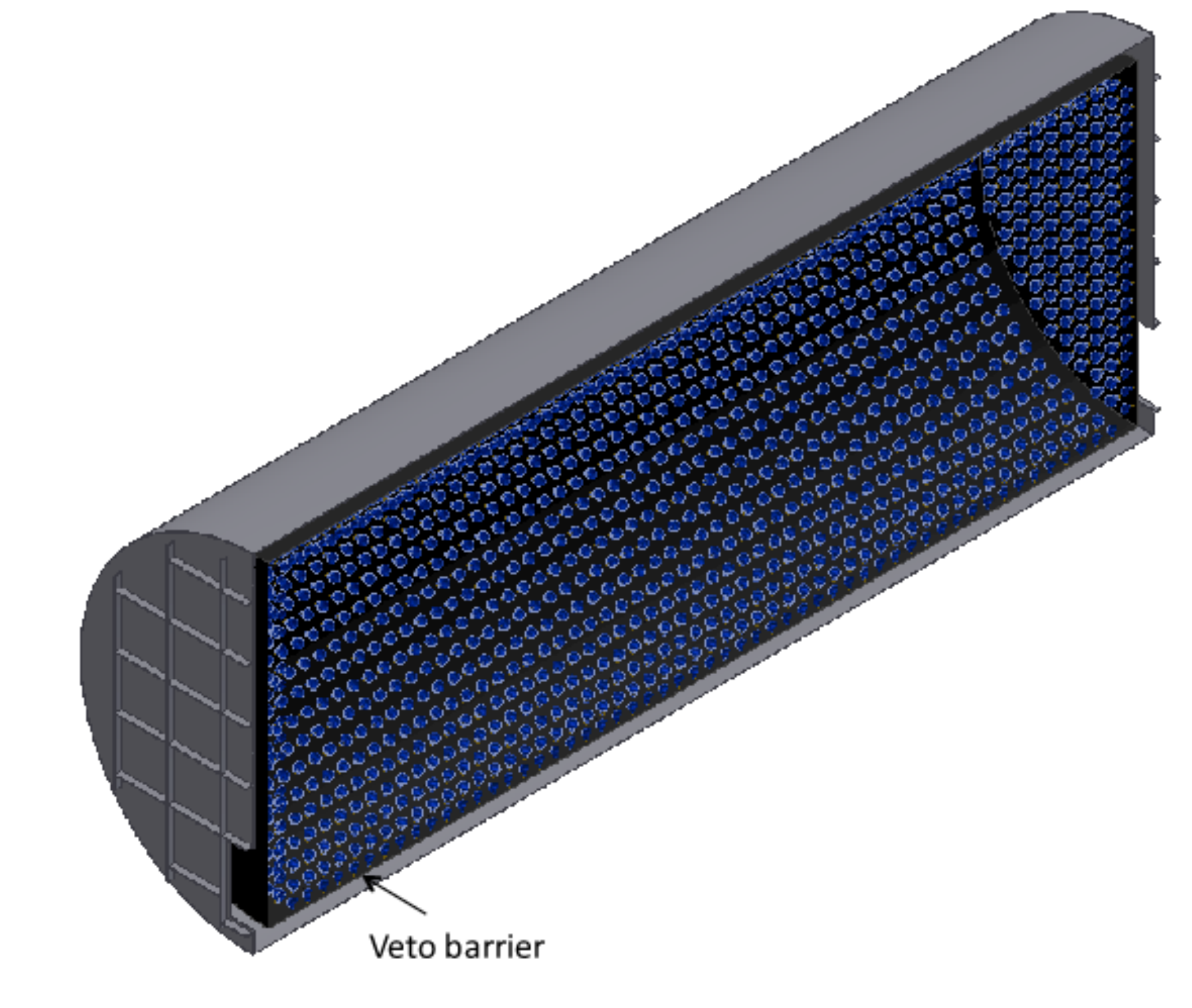}
\caption{
A cut-away schematic drawing of the OscSNS cylindrical detector tank, which is 8 m in diameter
and 20.5 m long. The outer 30 cm of the tank volume is a ``veto region'' that is optically isolated
from the ``detector region''. 
}
\label{detector_schematic}
\end{figure}

The OscSNS experiment will make use of this prodigious source of 
neutrinos.  The OscSNS detector will be centered at a location 60 meters from the SNS target, in the 
backward direction.  The cylindrical detector design is based upon the LSND and MiniBooNE detectors 
and will consist of an 800-ton tank of mineral oil (with a small concentration of b-PBD 
scintillator dissolved in the oil)
that is covered by approximately 3500 8-inch phototubes, yielding a photocathode coverage of 25\%. 
In addition, there will be approximately 300 8-inch phototubes for the veto region.
Figure~\ref{detector_target_schematic1} shows 
a schematic drawing of the suggested detector location in relation
to the SNS target hall, while Figure~\ref{detector_schematic} shows a cut-away schematic drawing of the OscSNS 
detector.

\begin{figure}
\centering
\includegraphics[width=16cm,angle=0,clip=true]{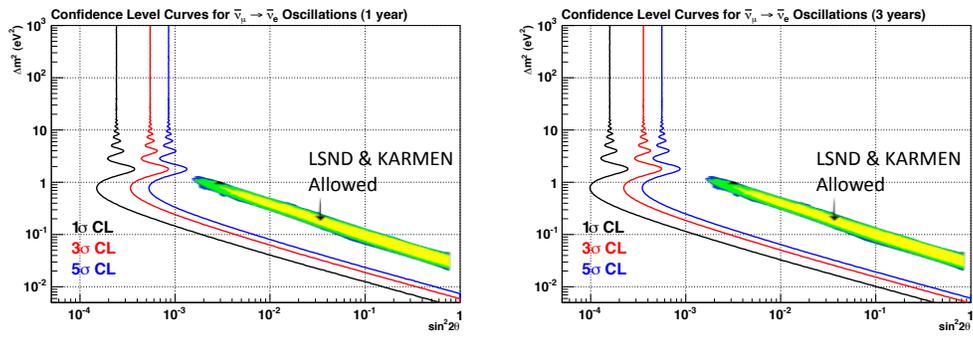}
\caption{
The OscSNS sensitivity curves for the simulated sensitivity to $\bar \nu_\mu \rightarrow \bar \nu_e$
oscillations after two (left) and six (right) calendar years of operation. Note that OscSNS has more than 5$\sigma$
sensitivity to the LSND allowed region after 2 years of data collection.
}
\label{app}
\end{figure}

\begin{figure}
\centering
\includegraphics[width=8cm,angle=90]{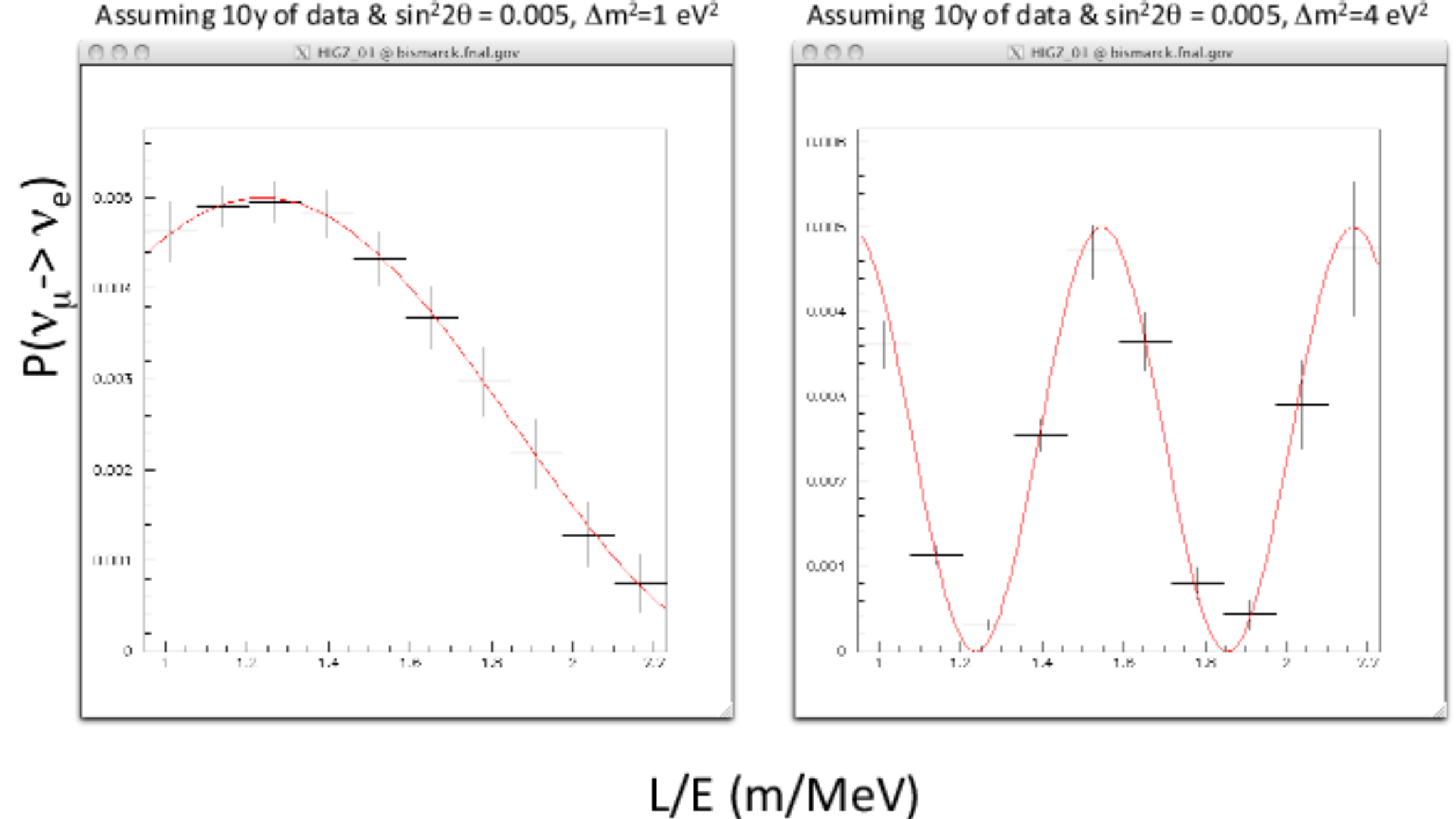}
\caption{The expected oscillation probability from $\bar \nu_e$ appearance as a
function of $L/E$ for
$\sin^22\theta = 0.005$ and $\Delta m^2 = 1$ eV$^2$ (left plot) and $\Delta m^2 = 4$ eV$^2$ (right plot).}
\label{loe_nuebar}
\end{figure}

This experiment will use the monoenergetic 29.8 MeV
$\nu_\mu$ to investigate the existence of light sterile neutrinos via the neutral-current 
reaction $\nu_\mu C \rightarrow \nu_\mu C^*$(15.11 MeV).  This reaction 
has the same cross section for all active 
neutrinos, but is zero for sterile neutrinos. An observed oscillation in this reaction is
direct evidence for sterile neutrinos.  OscSNS can also carry out an unique and decisive test of the 
LSND $\bar \nu_\mu \rightarrow \bar \nu_e$ appearance signal.
In addition, OscSNS can make a sensitive search for $\nu_e$ disappearance by searching for
oscillations in the reaction
$\nu_e C \rightarrow e^- N_{gs}$, where the $N_{gs}$ is identified by its beta decay. It is
important to note that all of the cross sections involved are known to two percent or better. 
Table~\ref{table:nums} summarizes the expected event sample sizes for the disappearance and appearance
oscillation searches, per calendar year, at the OscSNS.  Figure~\ref{app} shows the expected sensitivity for 
$\bar \nu_e$ appearance after two and six calendar years of run time. The LSND allowed region is fully
covered by more than 5$\sigma$.  The cylindrical design of the OscSNS detector allows for detection of 
oscillations as a function of $L/E$, as shown in Figure~\ref{loe_nuebar}.  Such an observation would
prove that any observed excess is due to short-baseline neutrino oscillations, and not due to a misunderstood 
background. The oscillation sensitivities are further improved by the construction of a near detector 
and by the planned construction of a second target station that is located at a longer neutrino baseline.

\begin{table}[ht]
\begin{center}
\begin{tabular}{|c|c|c|} \hline
 \multicolumn{1}{c}{Channel} &  \multicolumn{1}{c}{Background} & \multicolumn{1}{c}{Signal}  \\ \hline \hline
 \multicolumn{3}{c}{Disappearance Search}  \\ \hline
$\nu_\mu \ ^{12}C \rightarrow \nu_\mu \ ^{12}C^*$    &      & \\
$\nu_e \ ^{12}C \rightarrow \nu_e \ ^{12}C^*$    &      & \\
$\bar{\nu_\mu} \ ^{12}C \rightarrow \bar \nu_\mu \ ^{12}C^*$ & 1060 $\pm$ 36 & 3535 $\pm$ 182  \\ \hline
$\nu_\mu \ ^{12}C \rightarrow \nu_\mu \ ^{12}C^*$  & 224 $\pm$ 75 & 745 $\pm$ 42  \\ \hline
$\nu_e \ ^{12}C \rightarrow e^- \ ^{12}N_{gs}$  & 24 $\pm$ 13 & 2353 $\pm$ 123  \\ \hline
\multicolumn{3}{c}{Appearance Search} \\ \hline
$\bar \nu_\mu \rightarrow \bar \nu_e$: $\bar \nu_e \ ^{12}C \rightarrow e^+  \ ^{11}B \ n$  &  & \\
$\bar \nu_\mu \rightarrow \bar \nu_e$: $\bar \nu_e \ p \rightarrow e^+  \ n$          & 42 $\pm$ 5 & 120 $\pm$ 10 \\ \hline
$\nu_\mu \rightarrow \nue$: $\nue \ ^{12}C \rightarrow e^- \ ^{12}N_{gs}$  & 12 $\pm$ 3 & 3.5 $\pm$ 1.5 \\
\hline
\hline
\end{tabular}
\caption{Summary of per calendar year event rate predictions for a detector located at the SNS, centered at a distance of 60 meters from the interaction point, at $\sim$150
degrees in the backward direction from the proton beam.  The first column is the 
oscillation channel, the second column is the
expected intrinsic background, and the third column is the expected signal for appearance searches and the total
number of events for disappearance searches.  All event rates account for a 50\% detector efficiency, a 50\%
beam-on efficiency, a fiducial volume of 523 m$^3$, and are in units of expected events per calendar
year.  Appearance signal estimates assume a 0.26\% oscillation probability.}
\label{table:nums}
\end{center}
\end{table}

The SNS represents a unique opportunity to pursue a strong neutrino physics program in a cost-effective manner, 
as an intense flux of neutrinos from stopped $\pi^+$ and $\mu^+$ decay are produced during normal SNS operations. 
The OscSNS experiment would be able to prove whether
sterile neutrinos can explain the existing short-baseline anomalies.
The existence of light sterile neutrinos would be the first major extension of the Standard Model, and 
sterile neutrino properties
are central to dark matter, cosmology, astrophysics, and future neutrino research.

\bibliography{bibfile}

\end{document}